\pdfoutput=1
\documentclass[12pt]{article}
\usepackage{hyperref}
\usepackage{amsmath} 
\usepackage{amssymb}
\usepackage{graphicx}
\usepackage{gensymb}
\newcommand{\minus}{\scalebox{0.7}[1.0]{$-$}}
\def\be{\begin{equation}}
\def\ee{\end{equation}}
\def\ba{\begin{array}}
	\def\ea{\end{array}}
\def\bpm{\begin{pmatrix}}
	\def\epm{\end{pmatrix}}
\def\beqn{\begin{eqnarray}}
\def\eeqn{\end{eqnarray}}

\def\bt{\begin{tabular}}
	\def\et{\end{tabular}}
\def\bc{\begin{center}}
	\def\ec{\end{center}}

\begin{document} 
		\title{Minimal structure   for neutrino mass matrix} 
	\author {Manoj Kumar$^{1,3}$, Nikhila Awasthi$^2$, Monika Randhawa$^3$ and Manmohan Gupta$^2$
		\\
	{\small {\it 
		$^1$	Department of Physics, Akal university, Talwandi Sabo, Bathinda, 151302, India}}\\
		{\small {\it 
				$^2$ Department of Physics, Centre of Advanced Study, Panjab University,
		 Chandigarh, 160014, India.}}\\
			{\small {\it $^3$ University Institute of Engineering and Technology, Panjab
				University, Chandigarh, 160014, India.}}\\
	}
	
	\maketitle

	\begin{abstract}
	 	Taking clue from the minimal structure of  texture 4-zero
	hermitian mass matrices, which are very successful in accommodating quark mixing
	data,  we propose a form of   texture 2-zero complex
	symmetric neutrino mass matrix with only one phase parameter. This minimal mass matrix   not only accommodates the available
	neutrino oscillation data, but also makes interesting predictions for the unknown
	parameters like the lightest neutrino mass $m_{\nu_1}$ (for normal ordering(NO)), Jarlskog's
	rephasing invariant  $J_{CP}$, Dirac type CP violating phase $\delta_{CP}$  and effective neutrino mass  $\left< m_{ee} \right>$. We also explore the
	 correlations between the parameters of the model, that lead us to minimizing the parameters further.
	\end{abstract}
	\maketitle

\section{Introduction}
Determining the status of the CP symmetry in the lepton sector, discerning the hierarchy pattern of the neutrino masses, identifying the nature of massive neutrinos, i.e.  Dirac or Majorana and determining the absolute neutrino mass scale are among the highest priority goals of the researchers in neutrino physics. 
The measurement of CP violating phase and other oscillation parameters is important not only to understand the theory of neutrino oscillations, but 
also to  give  clues to reach the flavor theory and hence pave the way for building models of fermion masses and mixing. This, along with the increasingly precise determination of the lepton mixing parameters,  naturally leads us  to the next step, which is to probe the underlying physics  by the  construction   of the neutrino mass matrix followed by the detailed understanding of its structure giving rise to neutrino masses and  mixing.

In this regard, several approaches have
been developed to build the structure of neutrino
mass matrix. For example,  phenomenological ideas such as $\mu-\tau$ symmetry
\cite{mutau1,mutau2,mutau3,mutau4,mutau5,mutau6}, texture zeroes \cite{texzero1,texzero2,texzero3,texzero4,texzero5,texzero6,texzero7,
	texzero8,texzero9,texzero10,texzero11}, hybrid textures \cite{hybrid1,hybrid2,hybrid3,hybrid4,hybrid5}, vanishing minors \cite{minor1,minor2,minor3,minor4,minor5,minor6,minor7,minor8,minor9} etc. have been studied extensively from both bottom-up and top-down viewpoints.
It is to be noted that all these approaches insist on reducing
the number of free parameters in the mass matrix either
by invoking correlations among the mass matrix elements or restricting the mass matrix elements to take specific values so that
the mass matrix becomes predictive.

One of the ways of reducing the number of free parameters in fermion
mass matrices, is to assume some of their entries to be vanishing, i.e. to assume what
are commonly known as texture-zero ansatze \cite{weinberg,texfrz1,texfrz2}. Systematical and
complete analyses of all possible texture zeros have been carried out for both quark \cite{ludl-qk} and
lepton mass matrices \cite{ludl-lep1,ludl-lep2}.

In a recent work \cite{nikhila}, we formulated a minimal quark mass structure successful in reproducing the Cabibbo-Kobayashi-Maskawa (CKM) elements $\lvert V_{us} \rvert, \;\lvert V_{cb} \rvert, \lvert V_{ub} \rvert$ and angle $\beta$ of the unitarity triangle (UT).  Having found a minimal texture structure for quark mass matrices, it is natural to ask whether or not a similar texture for lepton mass matrices could reproduce the lepton mixing parameters within their experimental limits. The quark-lepton complimentarity, discussed by various authors \cite{QLC1,QLC2,QLC3} also hints towards some relation or parallelism in the two sectors. The possibility of a unique texture structure for quarks and leptons has also been explored in the literature \cite{unique1,unique2}. Therefore, inspired by the success of texture 4-zero quark mass matrices with only one phase, we would like to explore the implications of a similar structure of  mass matrices for the case of leptons. Assuming neutrinos to be Majorana particles, we propose a form of the neutrino mass matrix with only one phase parameter, that not only leads to a PMNS matrix consistent with experimental neutrino oscillation data, but also makes interesting predictions for unknown parameters like absolute neutrino mass, effective neutrino mass $\left< m_{ee} \right> $ and Majorana phases, $\eta_1$ and $\eta_2$.

The paper is organized as follows. In Section \ref{formulation}, we formulate a viable set of lepton mass matrices in the flavor basis and check for its viability against the latest oscillation data. In Section \ref{secpredic}, we discuss the predictions of our model for the unknown oscillation parameters. The correlations between the various parameters of our model have been discussed in the Section \ref{secpara}. In  Section \ref{secminimal}, we discuss the implications of simplifying our model by reducing the number of free parameters.
 Results pertaining to lepton mass matrices in non-flavor basis have been discussed in Section \ref{nonfl}.  Finally, Section \ref{sum} summarizes our conclusions.

By using polar decomposition theorem, though the number of parameters is
brought down to 18, it is still larger than the number of observables, therefore it
needs to be brought down further. In this context, following Weinberg [52] and
Fritzsch [39], the strategy has been to assume that mass matrices for fermions have
certain “textures” imposed on them by some underlying symmetries or these could
be purely phenomenological ansätze.

\section{A quark inspired minimal structure of lepton mass matrices}\label{formulation} 
It has been observed over the years that  texture 4-zero mass matrices, similar to the original Fritzsch ansatz \cite{texfrz1} have always been compatible with the quark and lepton mixing data  \cite{rrr,4zero1,4zero2,mmreview1,mmreview2,4zero3}.
The Fritzsch-Xing quark mass matrices (texture 2-zero for $M_u$ and texture 2-zero for $M_d$) are presented below,

\begin{equation}
M_u= \begin{pmatrix}
0&\mathbf{A}_u &0\\  \mathbf{A}^*_u &D_u& \mathbf{B}_u\\0&\mathbf{B}_u^*&C_u 
\end{pmatrix};~~
M_d= \begin{pmatrix}  
0&\mathbf{ A}_d&0\\\mathbf{ A}_d^*&D_d&\mathbf{B}_d\\0&\mathbf{B}_d^*&C_d 
\end{pmatrix},
\label{eq:ourMM} 
\end{equation}
where $\mathbf{A}_u=\lvert \mathbf{A}_u\rvert e^{i \alpha_u} = A_u e^{i \alpha_u}, \,\mathbf{A}_d=\lvert \mathbf{A}_d\rvert e^{i \alpha _d} = A_d e^{i \alpha _d}, \,\mathbf{B}_u=\lvert \mathbf{B}_u\rvert e^{i \beta_u} = B_u e^{i \beta_u}, \,\mathbf{B}_d=\lvert \mathbf{B}_d\rvert e^{i \beta_d} = B_d e^{i \beta_d}$, such that the combined texture of quark sector is  4-zero.
The predicting power of a mass matrix is maximal, when a minimal number of free parameters are introduced. Using simplicity and the requirement of maximal predictability as guiding principles, in Reference \cite{nikhila}, a set of texture 4-zero quark mass matrices with only one phase parameter were considered, which agreed excellently  with the latest experimental quark mixing data. For easy readability, we present these quark mass matrices   below,
\beqn
& &M_u= \begin{pmatrix}
0&A_ue^{i (\beta_u +\pi/4)}&0\\A_ue^{-i (\beta_u +\pi/4)}&D_u&B_ue^{i \beta_u}\\0&B_ue^{-i \beta_u}&C_u 
\end{pmatrix};  \nonumber \\
& & M_d= \begin{pmatrix}
0&A_de^{i (\beta_d -\pi/4)}&0\\A_de^{-i (\beta_d -\pi/4)}&D_d&B_de^{i \beta_d}\\0&B_de^{-i \beta_d}&C_d
\end{pmatrix}. 
\label{eq:6newMM}
\eeqn 

In the above set of mass matrices, the mixing matrix can be obtained in terms of only three free parameters, for example, $D_u, D_d$ and phase $\phi=(\beta_d - \beta_u)$. 
As mentioned in the introduction, there are various hints towards some relation or parallelism in the quark and lepton sector, therefore we consider a similar texture  structure in the lepton sector as well with only one phase and study its implications for the neutrino oscillations. Since the hierarchy of charged leptons is very strong, we can consider the charged lepton mass matrix to be diagonal, also considered by several authors \cite{flbasis1,flbasis2,flbasis3,flbasis4,flbasis5,flbasis6,flbasis7}, for example

\be 
M_l=\begin{pmatrix}
m_e &0&0\\
 0& m_{\mu}&0\\
 0&0&m_{\tau} 
\end{pmatrix} ;~~
M_{\nu}= \begin{pmatrix}
0&A_{\nu}&0\\A_{\nu}&D_{\nu}&B_{\nu}e^{i \phi}\\0&B_{\nu}e^{i \phi}&C_{\nu}
\end{pmatrix}.\label{eq:flmm}
\ee 

Here  $m_e, m_{\mu}$ and $m_{\tau}$ are the masses of the charged leptons $e, \mu$ and $\tau$  and the neutrino mass matrix elements $A_{\nu},\,B_{\nu},\,C_{\nu}$ and $D_{\nu}$ are real. It is pertinent to mention that  our work is based on the assumption that neutrinos are Majorana particles and hence the neutrino mass matrix is complex symmetric.  Further, the neutrino mass matrix
 contains only one phase, i.e. $\phi$, in contrast  to several such analyses  where more than one phase has been considered \cite{texzero1,texzero2,texzero3,texzero4,texzero5,texzero6,texzero7,
 texzero8,texzero9,texzero10,ludl-lep1,ludl-lep2,unique1,unique2,4zero1,4zero2,4zero3}.
  In this particular case, we are considering the NO of neutrinos, i.e. $m_{\nu_1}<m_{\nu_2} \ll m_{\nu_3}$. We diagonalize the mass matrices $M_l$ and $M_{\nu}$ exactly and construct the Pontecorvo-Maki-Nakagawa-Sakata (PMNS) matrix \cite{pmns1,pmns2,pmns3,pmns4} given as
  
\be
U_{PMNS}= V^{\dagger}_{l}V_{\nu}=
\begin{pmatrix}
U_{e1} & U_{e2} & U_{e3} \\
U_{\mu 1} & U_{\mu 2} & U_{\mu 3} \\
U_{\tau 1} & U_{\tau 2} & U_{\tau 3} 
\end{pmatrix},
\label{eq:PMNS}
\ee
where $V_l$ and $V_{\nu}$ are unitary matrices which diagonalize the lepton mass matrices $M_l$ and $M_{\nu}$, for example,

\beqn
V^{\dagger}_{l}M_lV_{l}=M^\textrm{diag}_l \equiv \textrm{Diag} ( m_e, m_{\mu}, m_{\tau}), \label{eq:diag1}\\
V^{\dagger}_{\nu}M_{\nu}V^*_{\nu}=M^\textrm{diag}_{\nu} \equiv \textrm{Diag} ( m_{\nu_1}, m_{\nu_{2}}, m_{\nu_{3}}),
\label{eq:diag2}
\eeqn

where $M^{diag}_{l(\nu)}$ are real and diagonal, while $V_{l}$ and $V_{\nu}$ are unitary $3 \times 3$ matrices. The charged lepton mass eigenvalues are denoted by $m_e, m_{\mu}$ and $m_{\tau}$, while $m_{\nu_1}, m_{\nu_2} \text{ and } m_{\nu_3}$ are the neutrino mass eigenvalues. The diagonalizing matrix $V_{l}$  for the charged lepton mass matrix is obviously a unit matrix in the flavor basis. Hence, $ U_{PMNS}=V_{\nu}$ and Eq.\eqref{eq:diag2} now becomes,

\be
U^{\dagger}_{PMNS} M_{\nu} U^*_{PMNS}= M^{diag}_{\nu}. 
\label{eq:mdiag} \ee

We obtain the matrix elements of $V_{\nu}$ in terms of the smallest neutrino mass $m_{\nu_1}$, mass squared differences, $\Delta m^2_{21}\text{ and } \Delta m^2_{31}$, (2,2) neutrino mass matrix element $D_{\nu}$ and phase $\phi$, however the expressions for these matrix elements are very long and can not be presented here. It may be mentioned that we have used the exact expressions for 
 $V_{\nu}$ in our analysis and no approximations have been used. 

Before going into the details of the analysis, we would like to mention some of the essentials pertaining to various inputs. 
Various past \cite{K2K,MINOS1,MINOS2,past1,past2,past3,Kamland} and ongoing experiments \cite{T2K,Daya,Reno,SKcollab,icecube,minos,sno,t2kdelta1,t2kdelta2,novadelta} have led to a very precise oscillation data. The mass squared differences and mixing angles, used in the present analysis at 1$\sigma$ and 3$\sigma$ confidence level (CL). are given in Table \ref{tab:data}  \cite{ivan,nufit}.
\begin{table}[h]
			\centering
			\renewcommand\arraystretch{2}
		\begin{tabular}{|c|c|c|}			
			\hline Parameters &  Best fit $\pm$1$\sigma$ & $3\sigma$ range\\		
		\hline $\theta_{12}/^\circ$ & $33.67^{+0.73}_{-0.71}$ & 31.61 $\to$ 35.94\\ 		
		$\theta_{23}/^\circ$&$42.3^{+1.1}_{-0.9}$ & 39.9 $\to $ 51.1\\		
		$\theta_{13}/^\circ$& $8.58^{+0.11}_{-0.11}$ & 8.23 $\to$ 8.91 \\	
		$\delta_{CP}/^\circ$& $232^{+39}_{-25} $& 139 $\to$   350\\		
		$\frac{\Delta m_{21}^2}{10^{-5}eV^2}$ & ${7.41}^{+0.21}_{-0.20}$  & 6.81 $\to$ 8.03 \\		
		$\frac{\lvert \Delta m_{31}^2 \rvert }{10^{-3}eV^2}$&  ${2.505}^{+0.024}_{-0.026}$ & 2.426 $\to$ 2.586 \\
		\hline
			\end{tabular}
\caption{Neutrino oscillation parameters as given by the global analyses, assuming  Normal Ordering (NO) \cite{ivan,nufit}
 .}  \label{tab:data}
\end{table} 

\begin{figure}[h]
\centering
\includegraphics[scale=0.85]{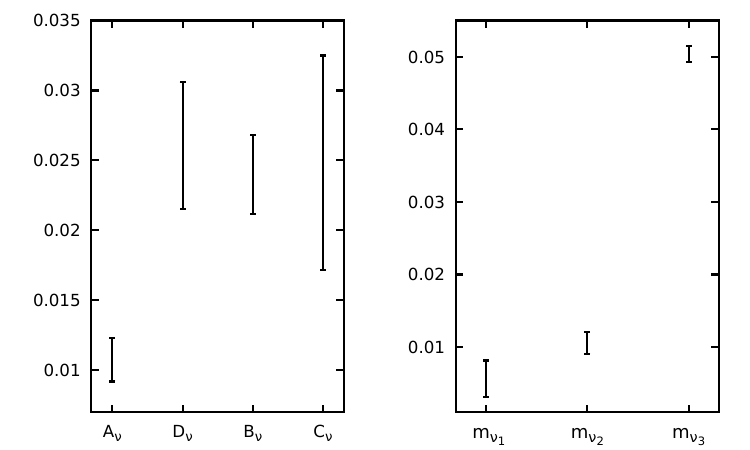}
\caption{Hierarchy exhibited by neutrino mass matrix elements (left panel) and  the neutrino masses (right panel) for the 2-zero neutrino mass matrix given in Eq. (\ref{eq:flmm}).}
\label{fig:hier}
\end{figure}

In the lepton sector, the mixing matrix $U_{PMNS}$  can be parameterized in terms of three mixing angles, one Dirac like $CP$ phase, $\delta_{CP}$, and two Majorana phases, $\eta_1$ and $\eta_2$, for example, {\small
\beqn
U_{PMNS}&=& \left( 
\arraycolsep=.6pt\ba{lll}
c_{12} c_{13}&s_{12} c_{13}&s_{13} e^{-i\delta_{CP}}\\
-s_{12} c_{23}-c_{12} s_{23} s_{13}e^{i\delta_{CP}}&c_{12} c_{23}-s_{12} s_{23} s_{13}e^{i\delta_{CP}}&s_{23} c_{13}\\
s_{12} s_{23}- c_{12} c_{23} s_{13} e^{i\delta_{CP}}&-c_{12} s_{23} -s_{12} c_{23} s_{13} e^{i\delta_{CP}}&c_{23} c_{13}\ea \right). Q , \\
\label{eq:PMNSparam}
Q&=& \left( \arraycolsep=.6pt \ba{lll}
e^{i \eta_1} &0 &0 \\
0&e^{i \eta_2} &0 \\
0 &0 &1\\
\ea  \right).
\eeqn }
Using Eqs.(\ref{eq:PMNS})$-$(\ref{eq:mdiag}), the PMNS matrix can be obtained in terms of   neutrino mass squared differences and free parameters $m_{\nu_1}$,  $D_{\nu}$ and $\phi$. For the analysis, we input the neutrino mass squared differences from Table \ref{tab:data} and scan their full ranges at 3$\sigma$, while the phase $\phi$ is scanned in the range $[-\pi,\pi]$. Thereafter, we calculate mixing angles $\theta_{12}, \theta_{23} \text{ and } \theta_{13}$. For free parameters, we keep the parameter sets in which the value of each observable (mixing angles given in Table \ref{tab:data}) is reproduced within the 3$\sigma$ interval of error-bars. We find that considering NO and the set of mass matrices given in Eq.(\ref{eq:flmm}), we are able to reproduce the mixing angles as given in Table \ref{tab:data} well within their experimental limits. The obtained ranges of free parameters for the NO case are given below in Eq. (\ref{eq:ranges}).
 
  \begin{eqnarray}    
m_{\nu_1} & = & (3.14-8.13) \times 10^{-3}eV, \nonumber \\
D_{\nu} & = & (0.022 - 0.031)eV , \nonumber\\
\phi & = & (\minus 14.6 - 14.6)\degree.
 \label{eq:ranges} 
\end{eqnarray}
The allowed ranges of the magnitudes of the matrix elements of $ M_{\nu}$ are
\begin{eqnarray}
M_{\nu}^r&=&\setlength{\arraycolsep}{10pt}\begin{pmatrix}
0 & 0.009 - 0.012& 0\\
0.009 - 0.012 & 0.022 - 0.031 & 0.021 - 0.027\\
0 & 0.021 - 0.027 & 0.017 - 0.032
\end{pmatrix}.
\end{eqnarray}
The absolute value of PMNS matrix is evaluated to be 
\begin{eqnarray}
|U_{PMNS}|&=&\setlength{\arraycolsep}{10pt}\begin{pmatrix}
 0.7998  - 0.8428 &  0.5178 -  0.5809 &  0.1431  - 0.1549\\
0.2729 -  0.5279  & 0.4921  - 0.5870 &  0.6345  - 0.7701\\
0.2785 -  0.5302 &  0.5649  - 0.6997  & 0.6205 -  0.7588
\end{pmatrix}. \quad \quad
\end{eqnarray}
We find that the obtained ranges of free parameters are quite narrow. Also, the neutrino mass matrix elements show a weak hierarchy. 
In Figure \ref{fig:hier},
we show the obtained ranges of neutrino masses as well as the corresponding ranges of neutrino mass matrix elements. The obtained ranges of neutrino masses are quite narrow and lean towards hierarchical neutrino masses, thus ruling out degenerate scenario. However, the hierarchy exhibited by neutrino mass matrix elements is very different. In particular, we observe that hierarchy between $A_{\nu}$ and other three elements is strong, while between $B_{\nu},\, D_{\nu}$ and $C_{\nu} $ is rather weak. 

It may be noted that the data at 3$\sigma$ allows $\phi$ to be zero as well. This means that even real symmetric neutrino mass matrix can accomodate the mixing angles given in Table \ref{tab:data}.
Further details of the parameter space spanned by  $ D_{\nu}, m_{\nu_1}\text{ and }\phi$ have been discussed   in Section \ref{secpara}.

\subsection{Predictions of minimal structure of lepton mass matrices}\label{secpredic}
The texture structure proposed by us is consistent with neutrino oscillation data as it reproduces mixing angles within their experimental ranges, however it would be desirable to ascertain if such a structure is predictive as well. 
In order to investigate the predictivity of our model, we calculate four observables, which are not experimentally measured yet, for example, lightest neutrino mass, effective neutrino mass and Majorana phases $\eta_1$ and $\eta_2$. We also calculate Dirac type $CP$ violating phase $\delta_{CP}$ measured recently by the long baseline experiments T2K  \cite{t2kdelta1,t2kdelta2}
and NO$\nu$A \cite{novadelta}. 

The effective neutrino mass $\left< m_{ee} \right>$ to be measured in neutrinoless double beta decay experiment is given as 

\be
\left< m_{ee} \right> = \lvert m_{\nu_1} U^2_{e1}+m_{\nu_2} U^2_{e2}+m_{\nu_3} U^2_{e3}
\rvert,
\ee

\noindent where  $U_{e1},U_{e2},U_{e3}$ correspond to the elements of $U_{PMNS}$ as given in Eq.(\ref{eq:PMNS}).
To evaluate the $CP$ violating phase $\delta_{CP}$, we make use of rephasing invariance of mass matrices. The Jarlskog's rephasing invariant parameter $J_{CP}$ is given as \cite{Jarlskog1,Jarlskog2}

\be
J_{CP} \sum_{\gamma , l} \epsilon_{\alpha \beta \gamma}\epsilon_{j k l } =\textrm{Im}[U_{\alpha j} U_{\beta k} U_{\alpha k}^* U_{\beta j} ^*]= J^{max}_{CP}\sin \delta_{CP}.
\ee
Using standard parameterization of the $U_{PMNS}$, $J^{max}_{CP}$ and hence the $CP$ violating phase $\delta_{CP}$ are given as 
\beqn J^{max}_{CP} &=& s_{12}s_{23}s_{13}c_{12}c_{23}c^2_{13}\,, \label{jmax}\\
\sin \delta_{CP} &=& \frac{J_{CP}}{J^{max}_{CP}}\,. \label{sindelta}
\eeqn
The Majorana phases $\eta_1$ and $\eta_2$ can be evaluated using the invariants \cite{branco}

\beqn
I_1=\rm{Im}(U^*_{e1}U_{e3}) \quad \text{and} \quad
I_2=\rm{Im}(U^*_{e2}U_{e3}), 
\label{eq:maj}
\eeqn
which can be solved to obtain the relationship between Majorana phases and the invariants as,

\be
\sin (\eta_1+\delta_{CP})= - \frac{I_1}{c_{12}c_{13}s_{13}}, \quad \text{and} \quad \sin (\eta_2+\delta_{CP})= - \frac{I_2}{c_{13}s_{12}s_{13}}.
\ee 
 
In Table \ref{tab:results}, we present the predictions of the mass matrices given in Eq.(\ref{eq:flmm}) for the lightest neutrino mass, $J^{max}_{CP}$, $\delta_{CP}$, $\left< m_{ee} \right>$, $\eta_1$ and $\eta_2$. \par
\begin{table}[h]
\centering
 \begin{tabular}{l}
 \hline \hline \\
$m_{\nu_1}=(3.14-8.13) \times 10^{-3}$eV   \\
$J^{max}_{CP}= (3.07 - 3.59) \times 10^{-2} $  \\ 
$\delta_{CP}=(0-90)\degree$ and  $(180-270)\degree$  \\
$\left< m_{ee} \right>=(0.0050-0.0081)$ eV   \\
$\eta_1=(\minus 143-143)\degree  $ \\
$\eta_2=(\minus 90-90)\degree$  
 \\
\hline \hline
 \end{tabular}
 \caption{Predictions of the minimal texture structure given in Eq.(\ref{eq:flmm}). }
 \label{tab:results}
\end{table}

On comparing with other recent analyses of mass matrices \cite{2109.04050v1,2106.15267,2107.12893,2106.07332}, we find that our results agree with most of these.
 For example, in Reference \cite{2109.04050v1},  the lightest neutrino mass $m_{\nu_1}$ is predicted between 2-8\,meV, while in  Reference \cite{2106.15267}, $m_{\nu_1} < 20\, \rm{meV}$ for NO.
 The strongest model independent constraint is provided from 
 the measurements of tritium $\beta-$decay by KATRIN experiment \cite{katrin}, which gives an upper bound of  0.45eV on the effective electron-neutrino mass 
 at 90\% CL.
 
The $CP$ violating Dirac phase $\delta_{CP}$ in lepton sector is yet to be measured precisely. The global fits to neutrino oscillation data \cite{ivan,nufit} give the best fit value for $\delta_{CP}$ to be  $232\degree$ in the case of NO of neutrino masses, however at 3$\sigma$ the range specified is quite broad and also includes $\delta_{CP}=180\degree$, i.e. CP conservation  as well.  The  phase $\delta_{CP}$ predicted by our model is within this range. 
Since as yet there are no clear cut conclusions on the value of $\delta_{CP}$, with even CP conservation  still allowed at a confidence level  of 1-2$\sigma$, our model's prediction in this regards should be taken seriously.
 \par

The Particle Data Group (PDG) \cite{pdg} provides the limit, $ J^{max}_{CP}= 0.0330\pm0.0006 (\pm0.0019)$ at  1$\sigma(3\sigma$CL), which is very much closer to the range predicted by our model. Further, the magnitude of $J_{CP}$ is predicted to be as large as 3.6\%, which suggests the possibility to measure leptonic $CP$ violation of this magnitude in the next-generation long-baseline neutrino oscillation experiments DUNE \cite{dune1,dune2} and HK \cite{hk1,hk2}, if the terrestrial matter effects could be taken care of. \par

For the case of effective mass $\left<m_{ee} \right>$, experimentally,  the strongest bound  comes from the KamLAND-
Zen experiment \cite{kamland-zen}, for example, $\left<m_{ee}\right >$ in the range of 0.028 - 0.122 eV (90\% CL). In Reference \cite{2109.14211v1}, the authors have studied the connection between dark matter and neutrinoless double beta decay in a scotogenic model with hybrid texture in the neutrino mass matrix. They observe lower limits on $\left< m_{ee} \right>$ which is compatible with the range calculated by us. In Reference \cite{2108.04066v1}, authors consider the implementation of U(1) $L_e-L_{\tau}$ gauge symmetry to study the neutrino phenomenology within the framework of type-(I+II) seesaw. They evaluate the maximum value of $\left<m_{ee} \right>$ to be around $0.025eV$. In Reference \cite{2109.04050v1}, the model specifically predicts the effective mass of neutrinoless double beta decay between 4-10 m$eV$. All these results are consistent with the range of $\left<m_{ee} \right>$ obtained by us.

It is imperative to note that our model is giving a restrictive range for $\eta_1$ and $\eta_2$. An experimental determination of Majorana phases would lead to ruling out several models of mass matrices. \par

\subsection{Parameter space spanned by free parameters of minimal structure of lepton mass matrices}\label{secpara}
\begin{figure}[h]
	\centering
	\includegraphics[width=.75\textwidth]{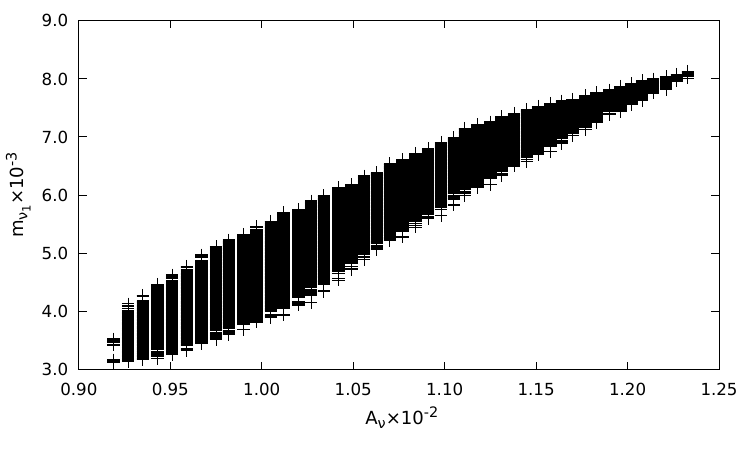}
	\caption{Variation of $A_{\nu}$ w.r.t. $m_{\nu_1}$ for the case of neutrinos following NO and having the structure of mass matrices as given in Eq.(\ref{eq:flmm}).}
	\label{fig:am}
\end{figure}
Exploring our model further, several important points emerge. 
We find that the allowed ranges of neutrino mass matrix elements, strongly affect the predictions of our model, particularly for the lightest neutrino mass, i.e. $m_{\nu_1}$. We find that $A_{\nu}$ is almost linearly correlated with $m_{\nu_1}$ as shown in Figure \ref{fig:am}. Thus a measurement of absolute neutrino masses will constrain our model. Other matrix elements, i.e $B_{\nu}, C_{\nu}$ and $D_{\nu}$ do not exhibit as strong a correlation with $m_{\nu_1}$. 
\begin{figure}[h]
\centering
\includegraphics[width=.75\textwidth]{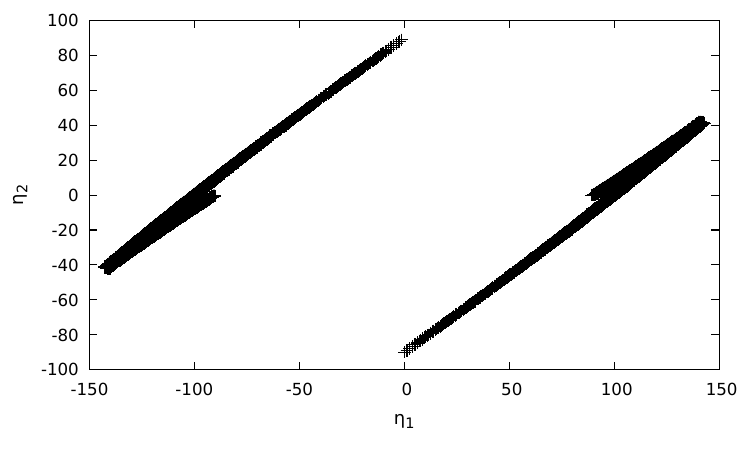}
\caption{Allowed region for Majorana phases, $\eta_1 $ and $\eta_2$, for the case of neutrinos following NO and having the structure of mass matrices as given in Eq.(\ref{eq:flmm}).}
\label{fig:eta1-2}
\end{figure}
In Figure \ref{fig:eta1-2}, we have plotted the predictions of lepton mass matrices given in Eq.(\ref{eq:flmm}) for Majorana phases $\eta_1$ and $\eta_2$.
The figure  indicates a strong correlation between $\eta_1$ and $\eta_2$.

 \begin{figure}[t]
\centering
\includegraphics[scale=0.9]{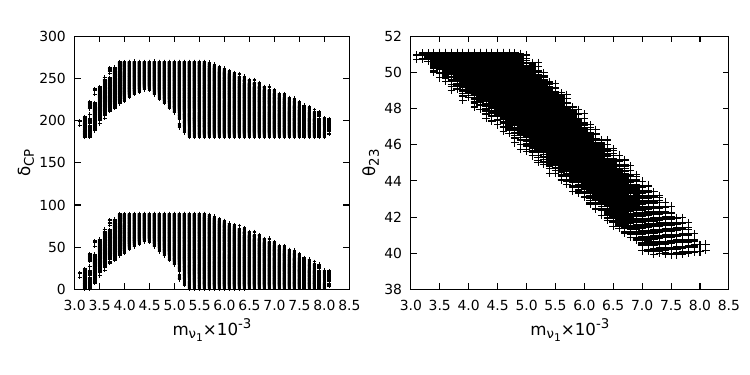}
\caption{Allowed region for lightest neutrino mass ($m_{\nu_1}$) and  $\delta_{CP}$ (left panel) and $m_{\nu_1}$ and  $\theta_{23}$ (right panel) for the case of neutrinos following NO and having the structure of lepton mass matrices as given in Eq.(\ref{eq:flmm}).}
\label{fig:mdelt23}
\end{figure}

In Figure \ref{fig:mdelt23} (left panel), we have plotted the predicted parameter space available to lightest neutrino mass $m_{\nu_1}$ and the Dirac $CP$ phase $\delta_{CP}$. These two parameters display a strong correlation, such that a discovery of $\delta_{CP}$ range will give strong constraints on the range of $m_{\nu_1}$. Also, from the right panel, we can infer that the precise measurement of $\theta_{23}$ will affect the predictions for $m_{\nu_1}$  as there is a significant correlation between these two. For lower values of  $\theta_{23}$, $m_{\nu_1}$ is pushed towards higher values in the allowed range given in the Table \ref{tab:data}.
Thus more precise data on three mixing angles may provide a sensitive test of this phenomenological scenario.

 A careful scrutiny of the left panel of Figure \ref{fig:hier} reveals that there is a  wide range of common values for the  neutrino mass matrix elements at (2,2), (2,3) and (3,3) positions i.e. $D_{\nu}$,  $B_{\nu}$ and $C_{\nu}$, for example $(2.16 - 2.68) \times 10^{-2}$eV. 
 This  inspires us to simplify our model further, by considering a neutrino mass matrix with matrix elements (2,2)=(2,3)=(3,3) in Eq. (\ref{eq:flmm}). In the next section, we study the phenomenological implications of this simplification. 

\subsection{Minimizing the parameters further}\label{secminimal}
Inspired by the common viable range of   neutrino mass matrix elements $D_{\nu}$, $B_{\nu}$ and $C_{\nu}$, we consider a neutrino mass matrix in the flavor basis, where $D_{\nu} =B_{\nu}= C_{\nu}$, as given below

\be 
M_l=\begin{pmatrix}
m_e &0&0\\
 0& m_{\mu}&0\\
 0&0&m_{\tau} 
\end{pmatrix} \text{ and }
M_{\nu}= \begin{pmatrix}
0&A_{\nu}&0\\A_{\nu}&D_{\nu}&D_{\nu}e^{i \phi}\\0&D_{\nu}e^{i \phi}&D_{\nu}
\end{pmatrix}. \label{eq:min}
\ee 
The above neutrino mass matrix is more predictive than the one given in Eq.(\ref{eq:flmm}), as there is only one free parameter, i.e. $m_{\nu_1}$.
The predictions for the above neutrino mass matrix are given  in
 Table \ref{tab:results1}.
 
\begin{figure}[h]
	\centering
	\includegraphics[scale=1.1]{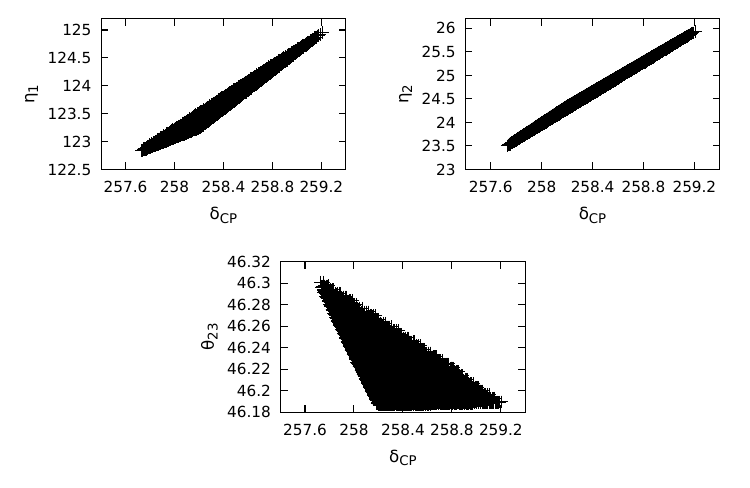}
	\caption{Correlations of $\delta_{CP}$ with $\eta_1$, $\eta_2$  and $\theta_{23}$ for the case of neutrinos following NO and having    mass matrix structure as given in Eq.(\ref{eq:min}).}
	\label{fig:corr}
\end{figure}

 \begin{table}[h]
 	\centering
 	\begin{tabular}{l}
 		\hline \hline \\
 		$m_{\nu_1}=(4.84-5.15) \times 10^{-3}$eV   \\
 		$J^{max}_{CP}=0.033 - 0.035  $\\
 		$\delta_{CP}=(257.6-259.2)\degree $  \\
 		$\left< m_{ee} \right>=(0.0065-0.0070)$ eV   \\
 		$\eta_1=(122-125)\degree  $ \\
 		$\eta_2=(23-26)\degree$  
 		\\
 		\hline \hline
 	\end{tabular}
 	\caption{Predictions of the minimal texture structure given in Eq.(\ref{eq:min}). }
 	\label{tab:results1}
 \end{table}
On comparing tables \ref{tab:results} and \ref{tab:results1}, we find that the above mass matrix predicts a much restrictive range of the unknown neutrino oscillation parameters. In particular, the smallest neutrino mass $m_{\nu_1}$ and Majorana phases $\eta_1$ and $\eta_2$ are much more constrained. 
In addition to the correlations shown in figures \ref{fig:am} to \ref{fig:mdelt23}, this minimalistic neutrino mass matrix shows very profound correlations of $\delta_{CP}$ with $\eta_1$, $\eta_2$ as well as with $\theta_{23}$,  as shown in Figure \ref{fig:corr}. It is clear that a precise determination of $\delta_{CP}$ will have significant implications for our model. In particular, a restrictive range of $\delta_{CP}$ will lead to our model predicting a very narrow range of  $\eta_1$ and $\eta_2$. Further, a precise measurement of $\theta_{23}$ will affect the predictions for  $\delta_{CP}$  as there is a significant correlation between these two. Thus more precise data on three mixing angles may provide a sensitive
test of this phenomenological scenario.

\section{A minimal structure of lepton mass matrices in non-flavor basis}\label{nonfl}

For the  sake of completion, we have carried out the entire analysis in the non-flavor basis as well, wherein the charged lepton mass matrix is also texture two zero type,  for example

\be 
M_l= \begin{pmatrix}
	0&A_l&0\\A_l&D_l&B_le^{i \phi}\\0&B_le^{i \phi}&C_l
\end{pmatrix} \text{ and }
M_{\nu}= \begin{pmatrix}
	0&A_{\nu}&0\\A_{\nu}&D_{\nu}&B_{\nu}e^{i \phi}\\0&B_{\nu}e^{i \phi}&C_{\nu}
\end{pmatrix}.
\label{eq:nonfl}
\ee

It is to be noted that keeping symmetry in mind, we have taken both Majorana neutrino mass matrix and charged lepton mass matrix as complex symmetric, although the charged leptons being Dirac particles, the lepton mass matrix may be taken  hermitian as well. Further, (2,3) elements of both neutrino mass matrix and charged lepton mass matrix contain the same phase $\phi$. Thus the above set of mass matrices contain fewer number of free parameters ($D_l, \, D_{\nu}, m_{\nu_1} \text{ and } \phi$) in comparison to several such analyses of texture 4-zero \cite{4zero1,4zero2,4zero3}, where more than one phase has been considered. Both $M_l$ and $M_{\nu}$ can be exactly diagonalized.

Considering the NO of neutrinos, we find that the above mass matrices successfully reproduce the mixing angles, given in Table \ref{tab:data}, within their 3$\sigma$ limits.  
The allowed ranges of the magnitudes of the matrix elements of $ M_{\nu}$ are
\begin{eqnarray}
M_{\nu}^r&=&\begin{pmatrix}
0 & 0.0096 - 0.0109& 0\\
0.0096 - 0.0109 & 0.0371 - 0.0397 & 0.0199 - 0.0217\\
0 &0.0199- 0.0217& 0.0044 - 0.0112
\end{pmatrix}.
\label{eq:nonflmnu}
\end{eqnarray}
We see that the hierarchy followed by neutrino mass matrix elements in the non-flavor case is different from the flavor case.  Unlike the  flavor basis case, the hierarchy between $C_{\nu}$ and $D_{\nu}$ is strong,  for example  $A_{\nu}\sim C_{\nu} < B_{\nu} \ll D_{\nu} $. 

The PMNS matrix is almost the same as for the flavor basis case and is given as
\begin{eqnarray}
|U_{PMNS}|=\begin{pmatrix}
 0.7999 -   0.8429 &   0.5178  -  0.5803  &  0.1431  -  0.1549\\
0.2612  -  0.3771  &  0.5492  -  0.6243  &  0.6956   - 0.7702\\
0.4060  -  0.5149 &   0.5661 -   0.6499  &  0.6204   - 0.7025
\end{pmatrix}. \quad \quad
\end{eqnarray}

Further, in the Table \ref{tab:nonflresults}, we present the predictions of these mass matrices for the lightest neutrino mass, Jarlskog's rephasing invariant parameter in the leptonic sector $J^{max}_{CP}$ and the corresponding Dirac like $CP$ violating phase $\delta_{CP}$, effective neutrino mass $\left< m_{ee} \right>$ and Majorana phases $\eta_1$ and $\eta_2$. \par
\begin{table}[h]
	\centering
	\begin{tabular}{ccc}
		\hline \hline \\ 
		$m_{\nu_1}=(0.97-2.94) \times 10^{-3}eV$ , &
		$J^{max}_{CP}=0.031 - 0.036$ ,&
		$\delta_{CP}=(35 - 85)\degree$ , \vspace{5pt} \\
		$\left< m_{ee} \right>=(3.45-4.68) \times 10^{-3} eV$ , &
		$\eta_1=(\minus 155 - \minus77 )\degree$ ,&
		$\eta_2=(\minus39 - 30)\degree$. \vspace{5pt} \\
		\hline \hline
	\end{tabular}
	\caption{Predictions of minimal texture structure for leptons considering NO in non-flavor basis, such that $M_l$ and $M_{\nu}$ has structure given in Eq.(\ref{eq:nonfl}). }
	\label{tab:nonflresults}
\end{table}

One can see that the predictions in the non-flavor basis are only slightly different from the flavor basis and are in tune with the latest analyses. It is pertinent to mention that although the set of mass matrices given in Eq. (\ref{eq:nonfl}) is able to reproduce the mixing angles, given in Table \ref{tab:data}, within their 3$\sigma$ limits, the value of $\theta_{23}$ leans towards higher octant. Furthermore, this set of  mass matrices is not viable, if all input parameters and constraints are taken at 1$\sigma$ CL, particularly it is not able to reproduce the mixing angle  $\theta_{23}$.

\par
We have carried out the entire analysis for both flavor as well as non-flavor basis, by also considering the IO of neutrinos, i.e. $m_{\nu_3} \ll m_{\nu_1} < m_{\nu_2} $. We find that in both cases, the given set of  mass matrices are not able to reproduce mixing data for IO, in particular, mixing angle $\theta_{23}$ cannot be reproduced.

\section{Summary and Conclusions}\label{sum}
Inspired by a minimal structure of quark mass matrices, we formulate texture 2-zero neutrino mass matrix with only one complex element, the charged lepton mass matrix being diagonal. We find that such a minimal neutrino mass matrix successfully reproduces the current neutrino mixing data. Assuming neutrinos following NO, we predict the numerical structure of charged lepton mass matrix, neutrino mass matrix and the PMNS matrix. The predictions of our set of mass matrices for lightest neutrino mass $m_{\nu_1}$, $J_{CP}$, $\delta_{CP}$ and $\left< m_{ee} \right>$ are found to be in tune with the latest analyses. In particular, the Majorana phases exhibit much constrained ranges and are strongly correlated, which might have useful implications for experimental searches related to these phases. We observe that the hierarchy between $A_{\nu}$ and other three elements is strong, while between $B_{\nu},\, D_{\nu}$ and $C_{\nu} $ is rather weak. The mass matrix element $A_{\nu}$ is observed to be almost linearly correlated with $m_{\nu_1}$ which shows that a measurement of absolute neutrino masses will constrain our model. The parameters $\delta_{CP}$ and $m_{\nu_1}$ display a strong correlation, such that a discovery of $\delta_{CP}$ range will give strong constraints on the range of $m_{\nu_1}$. We also infer that the precise measurement of $\theta_{23}$ will affect the predictions for $\delta_{CP}$ as there is a significant correlation between these two as well.\par

Most interestingly, we find that there is a  wide range of common values for the  neutrino mass matrix elements  $D_{\nu}$,  $B_{\nu}$ and $C_{\nu}$. Therefore, we also consider neutrino mass matrix with $D_{\nu}=B_{\nu}=C_{\nu}$, which is obviously more predictive  as there is only one free parameter, i.e.  $m_{\nu_1}$.  This minimalistic neutrino mass matrix shows  profound correlations of $\delta_{CP}$ with $\eta_1$ and $\eta_2$.

 We have carried out the entire analysis in the non-flavor basis as well, wherein the charged lepton mass matrix is also texture two zero type. One can see that the predictions in the non-flavor basis are only slightly different from the flavor basis.
For the sake of completeness, we have also analyzed the IO case and found that the given set of mass matrices are not able to accommodate IO oscillation data. 

\section*{Acknowledgment}
MK and MR would like to thank the Director, UIET,  Panjab University, Chandigarh  for providing facilities to work. NA and MG would like to thank the Chairman, Department of Physics, Panjab University, Chandigarh  for providing facilities to work.


\begin{thebibliography}{99}
	\bibitem{mutau1} T. Fukuyama, H. Nishiura, (1997) https://doi.org/10.48550/arXiv.hep-ph/9702253.
	\bibitem{mutau2} C. Lam, Phys. Lett. B \textbf{507}, 214 (2001).
	\bibitem{mutau3} P. F. Harrison, D. H. Perkins, W. G. Scott, 
	Phys. Lett. B \textbf{530}, 167 (2002).
	\bibitem{mutau4} S. King, C. Nishi, Phys. Lett. B \textbf{785}, 391 (2018).
	
	\bibitem{mutau5}Z.-z. Xing, Rept. Prog. Phys.  \textbf{86},  076201
	(2023).
	\bibitem{mutau6} P. Chakraborty, S. Roy,  Nucl. Phys. B \textbf{992}, 116252  (2023). 
	\bibitem{texzero1} H. Fritzsch, Z. Xing, Phys. Lett. B \textbf{353}, 114 (1995).
	
	\bibitem{texzero2} H. Fritzsch, Z. Xing, Prog. Part. Nucl. Phys. \textbf{45}, 1 (2000).
	\bibitem{texzero3}P. H. Frampton, S. L. Glashow and D. Marfatia,  Phys. Lett. B \textbf{536}, 79 (2002).
	\bibitem{texzero4} Z. Xing, H. Zhang, Phys. Lett. B \textbf{569}, 30 (2003).
	\bibitem{texzero5} M. Bando, S. Kaneko, M. Obara, M. Tanimoto, Prog. Theor. Phys. \textbf{112}, 533 (2004).
	\bibitem{texzero6} K. Matsuda, H. Nishiura, Phys. Rev. D \textbf{74}, 033014 (2006).
	\bibitem{texzero7} G. Ahuja \textit{et al.}, Phys. Rev. D \textbf{76}, 013006 (2007).
	\bibitem{texzero8} Z. Xing, Z. Zhao, Nucl. Phys. B \textbf{897}, 302 (2015).
	\bibitem{texzero9} M. Tanimoto and Tsutomu T. Yanagida, Prog. Theor. Exp. Phys.  \textbf{2016}, 043B03 (2016).
	\bibitem{texzero10} G. Ding, F. Joaquim, J. Lu, J. High Energy Phys. \textbf{2023}, 141 (2023).
	
	\bibitem{texzero11} R. Benavides \textit{et al.}, Phys. Rev. D \textbf{107}, 036008 (2023). 
	
	\bibitem{hybrid1} S. Kaneko, H. Sawanaka, M. Tanimoto, J. High Energy Phys. \textbf{08}, 073 (2005).
	\bibitem{hybrid2} S. Dev, S. Verma, S. Gupta, Phys. Lett. B \textbf{687}, 53 (2010).
	
	\bibitem{hybrid3} J.-Y. Liu, S. Zhou, Phys. Rev. D \textbf{87}, 093010 (2013). 
	\bibitem{hybrid4} W. Grimus, P. Ludl, J. Phys. G: Nucl. Part. Phys. \textbf{40}, 055003 (2013).
	
	\bibitem{hybrid5} S. Dev, D. Raj, Nucl. Phys. B \textbf{957}, 115081 (2020).
	
	\bibitem{minor1} E. Lashin, N. Chamoun, Phys. Rev. D \textbf{80}, 093004 (2009).
	
	
	\bibitem{minor2} S. Dev, S. Verma, S. Gupta, R. Gautam, Phys. Rev. D \textbf{81}, 053010 (2010). 
		\bibitem{minor3} S. Dev, S. Gupta, R. Gautam, L. Singh, Phys. Lett. B \textbf{706}, 168 (2011).
	\bibitem{minor4} T. Araki, J. Heeck, J. Kubo, J. High Energy Phys. \textbf{07}, 083 (2012).
	
	\bibitem{minor5} S. Dev, R. Gautam, L. Singh, Phys. Rev. D \textbf{89}, 013006 (2014).
	\bibitem{minor6} W. Wang, Phys. Lett. B \textbf{733}, 320 (2014).
	
	\bibitem{minor7} J. Liao, D. Marfatia, K. Whisnant, J. High Energy Phys. \textbf{09}, 013 (2014).
	\bibitem{minor8} I. Mazumder, R. Dutta, Phys. Rev. D \textbf{107}, 115023 (2023). 
	\bibitem{minor9} S. Dey, M. Patgiri, Phys. Rev. D \textbf{107}, 035012 (2023). 
	\bibitem{weinberg} S. Weinberg, Trans. New York Acad. Sci. \textbf{38}, 185 (1977).
	\bibitem{texfrz1} H. Fritzsch,   Phys. Lett. B \textbf{73}, 317 (1978).
	\bibitem{texfrz2} H. Fritzsch,   Nucl. Phys. B \textbf{155}, 189 (1979).
	
	\bibitem{ludl-qk} P. Ludl, W. Grimus, Phys. Lett. B \textbf{744}, 38 (2015).
	
	\bibitem{ludl-lep1} P. Ludl, W. Grimus, J. High Energy Phys. \textbf{1407}, 090 (2014).
	\bibitem{ludl-lep2} P. Ludl, W. Grimus, J. High Energy Phys. \textbf{1410}, 126 (2014).
	
	\bibitem{nikhila} N. Awasthi, M. Kumar, M. Randhawa, M. Gupta, Eur. Phys. J. C \textbf{82}, 653 (2022).
	
	\bibitem{QLC1} M. Raidal, Phys. Rev. Lett. \textbf{93}, 16 (2004). 
	\bibitem{QLC2} H. Minakata, A.Y. Smirnov, Phys. Rev. D \textbf{70}, 073009 (2004).
	\bibitem{QLC3} Z. Xing, Phys. Lett. B \textbf{679}, 111 (2009). 
	
	\bibitem{unique1} S. Sharma, P. Fakay, G. Ahuja, M. Gupta,  Int. J. Mod. Phys. A \textbf{29}, 1444005 (2014).
	\bibitem{unique2} M. Randhawa, V. Bhatnagar, P. Gill, M. Gupta, Phys. Rev. D \textbf{60}, 051301 (1999).
	

		\bibitem{rrr} P. Ramond, R. G. Roberts and G. G. Ross, Nucl. Phys. B \textbf{406}, 19 (1993).
	\bibitem{4zero1} Z. Xing, H. Zhang, J. Phys. G: Nucl. Part. Phys. \textbf{30}, 129 (2004). 
	\bibitem{4zero2} R. Verma \textit{et al.}, J. Phys. G: Nucl. Part. Phys. \textbf{37}, 075020 (2010). 
		\bibitem{mmreview1} M. Gupta, G. Ahuja, Int. J. Mod. Phys. A \textbf{26}, 2973 (2011).
	\bibitem{mmreview2} M. Gupta, G. Ahuja, Int. J. Mod. Phys. A \textbf{27}, 1230033 (2012).
	\bibitem{4zero3} S. Sharma, P. Fakay, G. Ahuja, M. Gupta, Phys. Rev. D \textbf{91}, 053004 (2015).
	
	\bibitem{flbasis1} D. Barreiros, R. Felipe, F. Joaquim, Phys. Rev. D \textbf{97}, 115016 (2018).
	\bibitem{flbasis2} M. Singh, Adv. High Energy Phys. \textbf{2018}, 2863184 (2018).
	\bibitem{flbasis3} Y. Kawamura, (2020) https://doi.org/10.48550/arXiv.2004.07664.
	\bibitem{flbasis4} Z. Xing, Phys. Rep. \textbf{854}, 1 (2020). 
	\bibitem{flbasis5} D. Zhang, Nucl. Phys. B \textbf{961}, 115260 (2020).
	\bibitem{flbasis6} H. Borgohain, D. Borah, J. Phys. G: Nucl. Part. Phys. \textbf{48}, 7 (2021).
	\bibitem{flbasis7} A. Adam, Prog. Theor. Exp. Phys. \textbf{2021}, 053B01 (2021).
	
	
	\bibitem{pmns1} B. Pontecorvo, Zh. Eksp. Theor. Fiz. (JETP) \textbf{33}, 549 (1957).
	\bibitem{pmns2} B. Pontecorvo, Zh. Eksp. Theor. Fiz. (JETP) \textbf{34}, 247 (1958).
	\bibitem{pmns3} B. Pontecorvo, Zh. Eksp. Theor. Fiz. (JETP) \textbf{53}, 1771 (1967). 
	\bibitem{pmns4} Z. Maki, M. Nakagawa, S. Sakata, Prog. Theor. Phys. \textbf{28}, 870 (1962).
	\bibitem{K2K} M. Ahn \textit{et al.} (K2K Collab.), Phys. Rev. D \textbf{74}, 072003 (2006).
	
	\bibitem{MINOS1} P. Adamson \textit{et al.} (MINOS Collab.), Phys. Rev. Lett. \textbf{108}, 191801 (2012).
	\bibitem{MINOS2} P. Adamson \textit{et al.} (MINOS Collab.), Phys. Rev. D \textbf{86}, 052007 (2012). 
	
	\bibitem{past1} G. Ranucci, Eur. Phys. J. A \textbf{52}, 79 (2016).
	\bibitem{past2} A. Santo, Int. J. Mod. Phys. A \textbf{16}, 4085 (2001).
	\bibitem{past3} M. Apollonio \textit{et al.} (CHOOZ Collab.), Phys. Lett. B \textbf{420}, 397 (1998).
	
	\bibitem{Kamland} S. Abe \textit{et al.} (KamLAND Collab.), Phys. Rev. Lett. \textbf{100}, 221803 (2008).
	
	\bibitem{T2K}K. Abe \textit{et al.} (T2K Collab.), Phys. Rev. D \textbf{103}, 112008 (2021)
	
	\bibitem{Daya} D. Adey \textit{et al.} (Daya Bay Collab.), Phys.Rev.Lett. \textbf{121}, 241805 (2018).
	
	\bibitem{Reno}C. Shin \textit{et al.} (RENO Collab.), PoS ICHEP2020, 177 (2021).
	
	\bibitem{SKcollab}K. Abe \textit{et al.} (Super-Kamiokande Collab.), Phys. Rev. D \textbf{97}, 072001 (2018). 
	
	\bibitem{icecube} M. Aartsen \textit{et al.} (IceCube Collab.), Phys.Rev.D \textbf{99}, 032007 (2019).
	
	
	\bibitem{minos} P. Adamson \textit{et al.} (MINOS Collab.), Phys. Rev. Lett. \textbf{112}, 191801 (2014).
	
	\bibitem{sno} B. Aharmim \textit{et al.} (SNO Collab.), Phys. Rev. C \textbf{88}, 025501 (2013).
	
	\bibitem{t2kdelta1} K. Abe \textit{et al.} (T2K Collab.), Nature \textbf{580}, 339 (2020). 
	\bibitem{t2kdelta2} K. Abe \textit{et al.} (T2K Collab.), Nature \textbf{583}, E16 (2020).
	
	\bibitem{novadelta} M. Acero \textit{et al.} (NO$\nu$A Collab.), Phys. Rev. D \textbf{106}, 032004 (2022).
	
	\bibitem{ivan} I. Esteban \textit{et al.},  J. High Energy Phys. \textbf{09}, 178 (2020).
	
	\bibitem{nufit} NuFIT v5.3 (2024), http://www.nu-fit.org/.
	
	\bibitem{Jarlskog1} C. Jarlskog, Phys. Rev. Lett. \textbf{55}, 1039 (1985).
	\bibitem{Jarlskog2} C. Jarlskog, Z. Phys. C \textbf{29}, 491 (1985).
	
	\bibitem{branco} G. Branco, M. Rebelo, Phys. Rev. D \textbf{79}, 013001 (2009).
	
	\bibitem{2109.04050v1} V. Mummidi, K. Patel, J. High Energy Phys. \textbf{2021}, 42 (2021).
	
	\bibitem{2106.15267} E. Valentino, S. Gariazzo, O. Mena, Phys. Rev. D \textbf{104}, 083504 (2021).
	
	\bibitem{2107.12893} E. Gonzalez, G. Kane, K. Nguyen, Phys. Rev. D \textbf{105}, 046019 (2022).
	
	\bibitem{2106.07332} M. Miskaoui, M. Loualidi, J. High Energy Phys. \textbf{2021}, 147 (2021).
	
	\bibitem{katrin} M. Aker \textit{et al.} (KATRIN Collab.), (2024) https://doi.org/10.48550/arXiv.2406.13516.
	
	\bibitem{pdg} R. Workman \textit{et al.} (Particle Data Group), Prog. Theor. Exp. Phys. \textbf{2022}, 083C01 (2022).
	
	\bibitem{dune1} B. Abi \textit{et al.} (DUNE Collab.), (2020) https://doi.org/10.48550/arXiv.2002.03005. 
	\bibitem{dune2} B. Abi \textit{et al.} (DUNE Collab.), (2021) https://doi.org/10.48550/arXiv.2103.04797. 
	\bibitem{hk1} K. Abe \textit{et al.} (Hyper-Kamiokande Collab.), Prog. Theor. Exp. Phys. \textbf{2015}, 053C02 (2015).
	\bibitem{hk2} K. Abe \textit{et al.} (Hyper-Kamiokande Collab.), Prog. Theor. Exp. Phys. \textbf{2018}, 063C01 (2018).
	
	\bibitem{kamland-zen} S. Abe \textit{et al.}, (2024) https://doi.org/10.48550/arXiv.2406.11438. 
	
	\bibitem{2109.14211v1} Ankush, M. Kashav, S. Verma, B.C. Chauhan, Phys. Lett. B \textbf{824}, 136796 (2022).
	
	\bibitem{2108.04066v1} M. Behera \textit{et al.}, (2021) https://doi.org/10.48550/arXiv.2108.04066.
	
	
\end{thebibliography}
\end{document}